\documentclass[prb,twocolumn,showpacs,citeautoscript,letterpaper]{revtex4}

\usepackage{amsmath}
\usepackage{amssymb}
\usepackage{graphicx}
\DeclareGraphicsExtensions{.pdf,.eps}
\usepackage{multirow}
\usepackage{dcolumn}
\usepackage{bm}
\usepackage{ulem}
\usepackage[usenames,dvipsnames]{color}

\begin{document}

\preprint{APS/123-QED}

\title{The structure, energy, and electronic states of vacancies in Ge nanocrystals}

\author{Kenneth Bayus}
\affiliation{Department of Material Science and Engineering, 
Cornell University, Ithaca, New York 14850}
\author{O.\ Paz}
\affiliation{Department of Physics and Astronomy,
Rutgers University, Piscataway, New Jersey 08854-8019}
\author{S.\ P.\ Beckman}
\email{sbeckman@iastate.edu}
\affiliation{Department of Material Science and Engineering,
Iowa State University, Ames, Iowa 50014}

\date{\today}

\begin{abstract}

The atomic structure, energy of formation, and electronic states 
of vacancies in H-passivated Ge nanocrystals are studied by 
density functional theory (DFT) methods.  The competition 
between quantum self-purification and the free surface relaxations 
is investigated.  
The free surfaces of crystals smaller than 2 nm distort the Jahn-Teller 
relaxation and enhance the reconstruction bonds.  
This increases 
the energy splitting of the quantum states and reduces the 
energy of formation to as low as 1 eV per defect in the 
smallest nanocrystals.  
In crystals larger than 2 nm the observed symmetry of the Jahn-Teller 
distortion matches the symmetry expected for bulk Ge crystals.
Near the nanocrystal's surface the vacancy is found to 
have an energy of formation no larger than 0.5 to 1.4 eV per 
defect, but a vacancy more than 0.7 nm inside the surface has 
an energy of formation that is the same as in bulk Ge.  
No evidence of the self-purification effect is observed; the 
dominant effect is the free surface relaxations, which allow for the 
enhanced reconstruction.  
From the evidence in this paper, it is predicted that for moderate
sized 
Ge nanocrystals a 
vacancy inside the crystal will behave bulk-like and not interact 
strongly with the surface, except when it is within 0.7 nm 
of the surface.  
\end{abstract}


\maketitle

\section{Introduction}

Germanium is a particularly attractive material for use in semiconducting
devices.
The charge carriers have a high mobility due to their low effective mass and
it is possible to achieve a high level of n- and p-type dopant
activation.\cite{yu2001,chui2003} The designers of 
microelectronics initially focused on Si instead of
Ge because Ge lacks a native oxide that can be used as a dielectric.  
%
%
Fortunately this limitation can be overcome by several techniques 
that have been developed 
within the last decade:
a thin Si overlayer can be grown over the Ge so that SiO$_{2}$ can be used as 
the dielectric,\cite{lee2001}
a Ge-oxynitride dielectric layer can be grown over the Ge,\cite{shang2002}
or a high-$\kappa$ dielectric crystal, such as ZrO$_{2}$, can be 
used in the device\cite{chui2002}.
These techniques allow for the development of Si/Ge heterostructure 
devices such as metal-oxide-semiconductor field effect transistors, 
MOSFETs.  The heterostructure MOSFET is primarily Si so existing 
fabrication technology can be used, but Ge is included as a buried 
channel between the source and drain to allow for high-speed 
conductivity.\cite{lee2005,shang2004,shang2006,cheng2004}

Before Ge can become an industrially important material, it must be possible 
to introduce and control a variety of dopant species in the 
crystal.\cite{queisser1998}  
Due to the initial challenge of finding a suitable dielectric material, 
there has been much
less effort in understanding Ge as compared to Si.  Subsequently much less
is known about the control of dopants in Ge than in Si.  
There exists a close relationship between impurity diffusion and 
self-diffusion; therefore, it is fundamentally important to understand 
self-diffusion to understand the control of impurity atoms.

Following Refs.\ [\onlinecite{bracht2000}] and [\onlinecite{ganster2009}], the
self-diffusion coefficient $D\left(T\right)$ is written as a sum of the
vacancy ($v$), interstitial ($i$), and direct-exchange ($ex$) diffusion
mechanisms,
\begin{multline}
D\left(T\right) =
  f_{v}\left(T\right)C_{v}^{eq}\left(T\right)D_{v}\left(T\right) + \\
  + f_{i}\left(T\right)C_{i}^{eq}\left(T\right)D_{i}\left(T\right) +
  D_{ex}\left(T\right)
\end{multline}
where $f_{\eta}\left(T\right)$ are correlation factors,
$C_{\eta}^{eq}\left(T\right)$ are the equilibrium concentrations of the intrinsic
defects, and $D_{\eta}\left(T\right)$ are the diffusion coefficients
corresponding to $\eta=v,i,ex$.
The concentrations and diffusion coefficients are expressed in terms of their
thermodynamic quantities,
\begin{gather}
C_{\eta}^{eq}\left(T\right)=
  \exp\left[\frac{\Delta S_{f}^{\eta}}{k_{B}}\right]
  \exp\left[-\frac{\Delta E_{f}^{\eta}}{k_{B}T}\right]\label{conc} \\
D_{\eta}\left(T\right)=
  K_{m}\exp\left[\frac{\Delta S_{m}^{\eta}}{k_{B}}\right]
  \exp\left[-\frac{\Delta E_{m}^{\eta}}{k_{B}T}\right]\label{diff}
\end{gather}
where $\Delta S_{f}^{\eta}$ and $\Delta E_{f}^{\eta}$ are the entropy and
energy of formation, $\Delta S_{m}^{\eta}$ and $\Delta E_{m}^{\eta}$ are the
entropy and energy of migration, $k_{B}$ is the Boltzmann constant, and $K_{m}$
is a constant prefactor.
The constant $K_{m}$ is independent of temperature and depends on the lattice
geometry and the vibrational frequencies.
The entropy terms include both the configurational and the vibrational
entropies.
The energy of formation is determined by the atomic bonding at the defect 
site and the energy of migration is determined by the energy of the saddle 
point configuration along the minimum energy transition path 
between stable atomic configurations.  

Direct exchange is the slowest diffusion mechanism.  Straining the lattice 
to allow the atoms to move past each other requires a prohibitively large 
amount of energy.  The principal diffusion pathways involve either vacancy 
or interstitial assisted migration. The work presented here focuses on these 
intrinsic point defects.  
Based on the energies listed in TABLE~\ref{libtable} the
self-diffusion coefficient in Si is controlled by a 
self-interstitial kick-out mechanism at high temperatures
($T>900^{\circ}$~C) and vacancy-mediated diffusion at lower 
temperatures.~\cite{shimizu2007} 

In Ge, vacancy-assisted diffusion is the primary mode.~\cite{silvestry2006}
Although the migration barrier for Ge vacancies and self-interstitials
is roughly equivalent, the energy of formation for interstitial Ge
atoms is approximately 1~eV greater than vacancies, whereas in Si the 
energy to create vacancies and interstitials is equivalent.  
In Ge self-interstitial atoms will only be formed at very high
temperatures or after highly energetic processes such as
irradiation.\cite{BrachtPRL2009} Therefore, vacancies in Ge are
substantially more influential than interstitial atoms for assisting
diffusion under thermal equilibrium, as compared to Si.

\begin{table}
\caption{\label{libtable}
The energies of formation and migration for vacancies and interstitial defects
in Si and Ge.}
\begin{ruledtabular}
\begin{tabular}{ccccc}
Element & $\Delta E_{f}^{v}$          & $\Delta E_{f}^{i}$ 
        & $\Delta E_{m}^{v}$          & $\Delta E_{m}^{i}$          \\
\hline
\multirow{4}{*}{Si}
        & 3.1--3.6\cite{shimizu2007}  & 3.2\cite{zhu1996}
        & 0.4--1.40\cite{shimizu2007} & 0.45\cite{sahli2005}        \\
        & 3.7\cite{centoni2005}       & 3.31--3.84\cite{leung1999}
        & 0.43--0.49\cite{ganster2009}& 0.84\cite{ganster2009}      \\
        & 3.49\cite{antonelli1998}    & 3.27\cite{wang1991}
        &                             &                             \\
        & 3.53\cite{wright2006}       &
        &                             &                             \\
\hline
\multirow{4}{*}{Ge}
        & 2.3\cite{fazzio2000}        & 2.29\cite{dasilva2000}
        & 0.7\cite{uberuaga2002}      & 0.5\cite{CarvalhoPRL2007}   \\
        & 1.7--2.0\cite{hwang1968}    & 2.3--4.1\cite{chroneos2008}
        & 0.36--0.7\cite{Pinto2006498}&                             \\
        & 2.4\cite{uberuaga2002}      & 3.55\cite{Moreira2004}
        &                             &                             \\
        & 2.6\cite{Pinto2006498}      & 3.50\cite{Vanhellemont2007}
        &                             &                             \\
        & 2.56\cite{Vanhellemont2007} & 
        &                             &                             \\
\end{tabular}
\end{ruledtabular}
\end{table}
 
The dominance of vacancies-assisted diffusion is observed 
experimentally, both for
self-diffusion\cite{werner1985} and impurity-atom
diffusion\cite{silvestry2006,bracht2007,bracht1991}.
The interaction between vacancies and impurity atoms is complicated. It is 
believed that vacancies and impurities form mobile defect 
pairs.~\cite{brotzmann2008,chroneos2008,chroneos20082}  This 
defect pair can become 
pinned when a second impurity, such as C, joins the 
complex.~\cite{chroneos2008}  
In addition to forming complexes, the vacancies and impurities often carry a 
charge.~\cite{fazzio2000,brotzmann2008,chroneos2008}  It is likely that 
an isolated vacancy in bulk Ge is charged $-$2.\cite{brotzmann2008}  
In the work presented here only isolated, charge neutral, impurities are 
investigated, which is consistent with the nanoscale context of this 
study.  

To improve the engineering control of material properties and increase 
device efficiency, it is desirable to move from bulk to nanoscale structures.
There are many examples of situations where nanostructured Ge offers benefits.
The use of Ge nanocrystals as the floating gate of MOS memory devices results
in a dramatic shift in the threshold-voltage, improved switching
characteristics, and decreased leakage current.\cite{das2009,ma2008}
Ge films with nanostructured surfaces offer the ability to tune the optical
properties of thin films.\cite{sato1995}
Ge nanowires are considered for use as MOSFETs.\cite{zhang2008,jiang2008}

Ultimately the nanostructures used to create devices
need to be tailored by controlling their size, surfaces, and dopants.
With respect to introducing dopants, the electronic properties of
nanostructures are believed to be sensitive to 
the relative position of the impurities in the structure.
The mean free path of charge carriers within nanowires depends
strongly on the radial dopant profile.\cite{markussen2007} 
This will influence the conductivity.  
In addition to the challenge of selectively incorporating the dopant atoms 
into the nanostructure, the impurity distribution must be maintained for the 
lifetime of the device. 

At the quantum scale the primary difference between a bulk crystal and 
a nanocrystal is the interaction of the wave function with the 
surfaces.  
As the size of a structure decreases, the 
crystal's translational symmetry ceases to be meaningful.  
The electronic band-structure that is nominally a function of the quantum
number $\mathbf{k}$ is projected onto the $\Gamma$ point in the center of
the Brillouin zone.  
The crystal's energy bands become discrete quantum energy states. 
Whereas in bulk the wave function is distributed across the entire crystal 
as Bloch waves, 
$u_{\mathbf{k}}\left(\mathbf{r}\right)e^{i\mathbf{k}.\mathbf{r}}$, 
in nanocrystals the wave function is confined by the surfaces.  
The size of the nanostructure directly impacts the 
energy states, analogous to the elementary particle-in-a-box problem.  
Consider for example a [110] Ge nanowire.  When the wire diameter is 
sufficiently small the crystal's translational symmetry is only 
meaningful in the [110] 
direction and the bulk Ge states are projected along 
the $\mathbf{k}=\left[110\right]$ direction in k-space.  This projection 
transforms 
Ge from an indirect to direct band gap material.\cite{beckman2006}  
The confinement is predicted to distort the shape of the energy dispersion for 
wires with diameters as large as 2 nm.  The energy bands of nanowires 
with diameter greater than 2 nm are found to undergo a rigid shift, even for 
wires as large as 5 nm.\cite{beckman2006}

Fundamentally, there are two effects that differentiate the behavior of defects in 
nanostructures from bulk: quantum confinement and free surfaces.
Dalpian \cite{dalpian2006} claims that the confinement of the defect's wave function results 
in the so-called {\it self-purification} effect that increases the 
defect's $\Delta E_{f}^{\eta}$. In the case of dopant species, 
this increase hinders the incorporation of dopant atoms into the 
nanostructures.  This is a controversial subject and worthwhile 
investigating.\cite{du2008,dalpian2008,li2008} In the present calculations 
evidence of self-purification will be sought.  

The free surfaces allow the nanostructure to expand or contract to reduce the 
strain energy surrounding the defect.  From an energetics perspective the 
self-purification effect and the free surfaces compete with one another.  The 
self-purification increases the energy and the free surfaces decrease the energy. 
From a kinetics perspective they complement one another because it is likely 
that the surfaces will getter impurities out of the nanostructure.  
In the case of Si nanocrystals it is observed that the relative energy to
introduce vacancies decreases as the nanocrystal's size decreases.
This indicates that energetically the free surfaces dominate the
self-purification effect.\cite{beckman2007}
As the vacancy is moved toward the surface the energy further decreases and
when the vacancy is within 0.6 \AA\ of the surface it becomes unstable and is
spontaneously moved to the surface of the crystal.\cite{beckman2007,beckunpub}

In this paper the structure and energies of vacancies in Ge nanocrystals are
examined as a function of the nanocrystal's size and the position of the
vacancy in the crystal.
Because the energies, $\Delta E_{f}^{\eta}$, depend on the size of the crystal and the 
position within the crystal, the concentrations and diffusivities, 
from EQ.\ \ref{conc} and \ref{diff}, also 
depend on the size and position.
Self-diffusion within nanostructures is not a simple matter that can be
easily described by a single coefficient.  
The work here is a first step toward building a comprehensive model.

Following this introduction, in Sec.\ \ref{methods}, the methods used will be
presented, including a discussion of the computational approach and the
details of the nanocrystals' morphology.
The results from these calculations will be presented and discussed in Sec.\
\ref{results}.
A concluding summary will be presented in Sec.\ \ref{conclusion}.

\section{Methods\label{methods}}

\subsection{Computational approach}

The calculations are performed within the framework of the density-functional
theory\cite{DFT} (DFT), using the local-density approximation\cite{LDA} (LDA)
for the exchange-correlation functional, as it is implemented in the
\textsc{Siesta} computational package.\cite{Siesta}
The electrons in the core atomic region are substituted by norm-conserving
pseudopotentials of the Troullier-Martins type\cite{TroullierMartins}, and
the valence charge is represented by a set of atom-centered basis functions.
In \textsc{Siesta} these functions correspond to numerical atomic orbitals of
strictly finite range, a particular choice that is specially suited to treat 
isolated systems.

All calculations are carried out using a double-$\zeta$ plus polarization
orbitals (DZP) basis set.
The cutoff radii of the basis functions are optimized for bulk germanium in
the diamond structure, following the method proposed in Ref.\
[\onlinecite{BasisSets}], where a fictitious external pressure of 0.2 GPa is
employed on the free atom.
This basis size and radius lengths are proven to give a good balance between
the computational accuracy and cost.
A theoretical lattice parameter of $a_0=5.64$ \AA\ and a bulk modulus of
$B_0=80.0$ GPa, are obtained from the fitting to the Murnaghan equation of
state. \cite{Murnaghan}
Both values are in good agreement with the structural and elastic properties
from experiments\cite{LandoltBornstein} ($\tilde{a}_0=5.66$ \AA\ and
$\tilde{B}_0=75.8$ GPa), given the fact that the LDA tends to underestimate
lattice constants by a 1--3\%, but also to overestimate bulk moduli with
errors ranging from 5 up to 20\%.\cite{HamannPRL96}

The theoretical method employs periodic boundary conditions.  The 
nanocrystals are placed inside the supercell surrounded by a buffer of 
empty space.  
The size of the atomic clusters ranges from 44 to 244 Ge atoms and the vacuum
region is chosen to be large enough as to avoid any interaction between their
periodic replicas.
A kinetic energy cutoff of 250 Rydberg is chosen for the real-space 
integrations involving the Hartree and the exchange-correlation 
contributions to the self-consistent potential.
In this respect, a stringent criterion is employed in the convergence 
of the density matrix and total energy.  All atomic coordinates are then 
relaxed according to a conjugate-gradient minimization algorithm, until the 
maximum residual forces are below 0.02 eV/\AA.

\begin{figure*}
\includegraphics[clip=true,width=\textwidth]{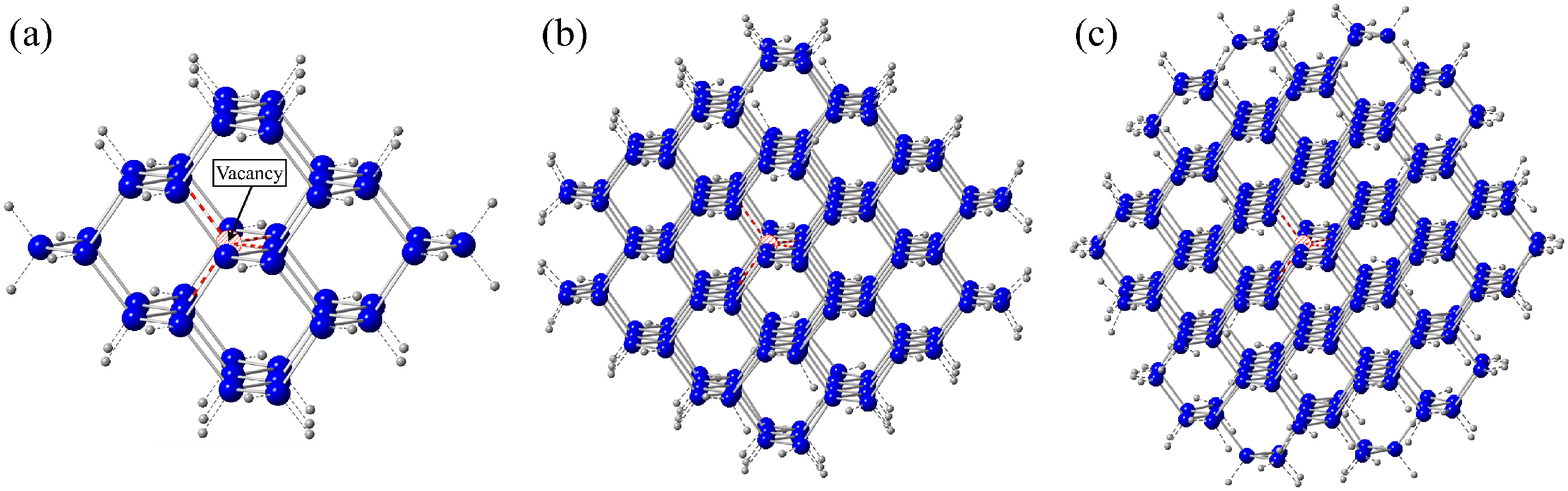}
\caption{\label{nanocrystals}
(color online)
Ge nanocrystal morphology for (a) Ge$_{44}$H$_{42}$, (b) Ge$_{130}$H$_{98}$, and (c) Ge$_{244}$H$_{158}$. 
The corresponding sizes are 1.02, 1.70 and 2.20 nm respectively.
A central vacancy is depicted as a hatched atom surrounded by four missing bonds (broken lines), all in red.
Ge atoms are colored in blue; saturating H atoms are in light grey.
}
\end{figure*}

\subsection{Nanocrystal morphology}

The nanocrystal geometries used in these studies are hydrogen passivated,
bond-centered crystals.
Experimental nanostructures frequently have amorphous, glassy, or polymeric
coatings that result from the method of crystal growth.
It is possible to treat the surfaces to reduce or remove these, although it is
uncommon experimentally to work with bare surfaces.
The nanocrystals investigated here have their surfaces passivated with
a extremely ``soft" H pseudopotential. 
Surface passivation removes surface 
states and allows the 
competition between the self-purification effect and the free surfaces 
to be studied without considering the complicated surface chemistry.  

Surface Ge atoms are identified and the dangling bonds of the Ge atoms are
capped with H.
Any Ge atom that is found to have three dangling bonds is replaced by a single
H atom.
The resulting nanocrystals are shown in FIG.\ \ref{nanocrystals}.
The surface morphologies are examined and crystals that are highly 
faceted are excluded.  Only nanocrystals with near spherical geometry are 
studied.

\section{Results and Discussion\label{results}}

\subsection{Atomic structure and defect states\label{asds}}

\begin{table}
\caption{\label{JTatom}
The local bond lengths at a vacancy site in crystals with 
diameters 1.02 and 2.20 nm.  The 
segment labels reference the tetrahedral structure in 
FIG.~\ref{tet}.
Using the DFT-LDA the theoretical bond length in bulk Ge is 2.44 \AA, which corresponds to segment lengths of 3.99 \AA\ before atomic relaxation.}
\begin{ruledtabular}
\begin{tabular}{ccc}
Segment & $D=1.02$ nm & $D=2.20$ nm \\
\hline
$\overline{\text{AB}}$ &2.55 &2.90 \\
$\overline{\text{AC}}$ &3.57 &3.47 \\
$\overline{\text{AD}}$ &3.57 &3.47 \\
$\overline{\text{BC}}$ &3.79 &3.46 \\
$\overline{\text{BD}}$ &3.79 &3.46 \\
$\overline{\text{CD}}$ &2.50 &2.90 \\
\end{tabular}
\end{ruledtabular}
\end{table}

\begin{figure}
\begin{center}
\includegraphics[width=0.75\columnwidth]{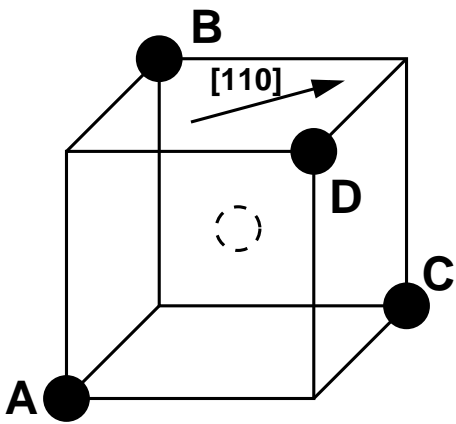}
\caption{The reference geometry of the atomic structure at the vacancy 
site.  The dashed circle is the vacancy and the solid circles are the 
nearest neighbor Ge atoms that form a tetrahedron.  The tetrahedron, without 
atomic motion has $T_{d}$ symmetry.  The calculated bond lengths for various 
sized nanocrystals are shown in TABLE \ref{JTatom}.}
\label{tet}
\end{center}
\end{figure}

\begin{figure}
\includegraphics[width=\columnwidth]{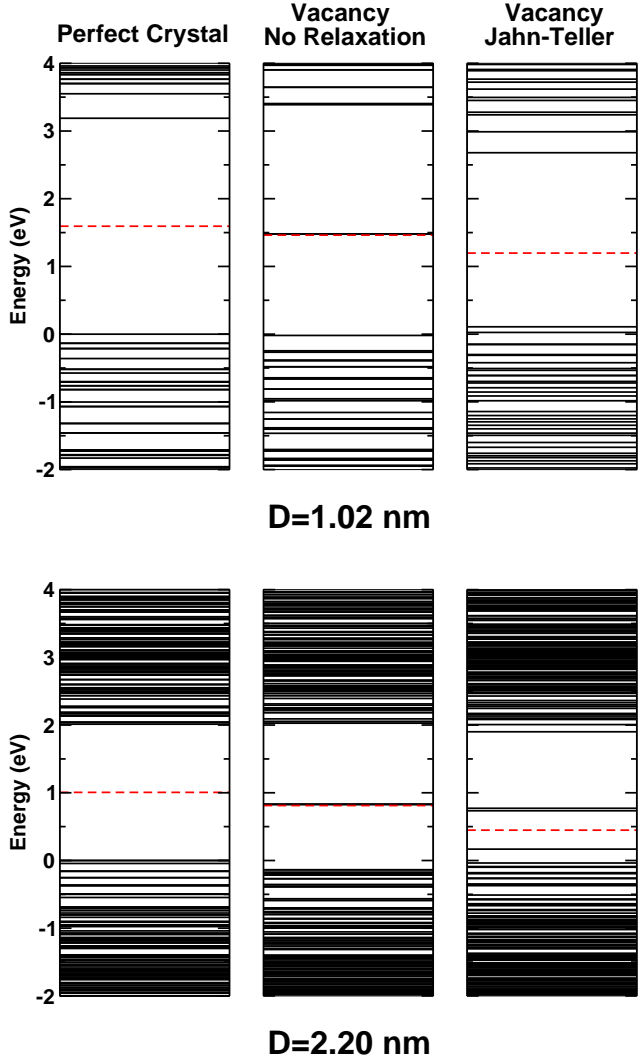}
\caption{
The top frame shows the energy levels for a crystal with 1.02 nm diameter 
and the bottom frame shows a 2.20 nm crystal. 
The red dashed line is the Fermi energy.  The band alignment between the 
diagrams is arbitrary.  
The crystals in the left column are perfect and contain no vacancy.  
The crystals in the center column have vacancies, but the structure 
has not been allowed to relax.  The states at the Fermi level are 
three-fold degenerate.  
The right column shows the states after the Jahn-Teller distortion is 
completed, which allows the atomic structure to break the symmetry and 
lift the degeneracy of the gap state.  
}
\label{statediagram}
\end{figure}

The local atomic structure at the vacancy site is directly related to the 
electronic states introduced to the gap from the broken bonds.  
The atomic structure of the vacancy in the 1.02 and 2.20 nm nanocrystals is 
given in TABLE \ref{JTatom}, using as reference the ABCD indices shown in 
FIG.~\ref{tet}.  
The associated electronic states are diagrammed 
in FIG.~\ref{statediagram}.  The left column of FIG.~\ref{statediagram} 
shows the states for a nanocrystal with no vacancy. The band gap for 
the 1.02 nm crystal is 3.13 eV and the gap for the 2.20 nm crystal 
is 2.0 eV.  

An undistorted vacancy has $T_{d}$ symmetry.  There are three 
degenerate states in the gap, belonging to the $t_{2}$ representation, 
associated with this structure.  
These 
are shown in the middle 
column of FIG.~\ref{statediagram}.  There are two electrons localized at 
the defect so the states are partially occupied.  It is the partial occupancy 
of the degenerate energy levels that allows the defect to undergo a spontaneous 
symmetry breaking that reduces the degeneracy and lowers the 
electronic energy.  In bulk crystals the Jahn-Teller distortion produces a 
$D_{2d}$ symmetrized structure with the fully-occupied state belonging to 
the $b_{2}$ representation and the doubly-degenerate, empty state having 
the $e$ representation.\cite{watkins1986,serdar2001,beckman2007} 

Here, as in the case of a vacancy in a Si nanocrystal\cite{beckman2007}, the 
symmetry of the structure approaches $D_{2d}$ but due to the 
surfaces there is additional distortion.  
In the 
case of the 1.02 nm crystal, the symmetry of the vacancy structure 
is $C_{s}$.  For the 2.20 nm crystal the symmetry is essentially $D_{2d}$ 
but a minuscule 0.01 nm distortion of the bonds lowers the 
symmetry, {\it i.e.}, if 
$\overline{\text{AC}}$=$\overline{\text{AD}}$=$\overline{\text{BC}}$=$\overline{\text{BD}}$ then the 
symmetry would be $D_{2d}$.
From TABLE \ref{JTatom} it is determined that the Jahn-Teller 
distortion in the 1.02 nm crystal is approximately 10\% larger than 
that in the 2.20 nm crystal and as a result the defect states undergo 
a larger energy split, over 2.7 eV, which almost 
pushes the states out of the gap.  In the 2.20 nm crystal the 
splitting is much 
smaller, around 0.75 eV. 

\subsection{Crystal size\label{cryssize}}

\begin{figure}
\includegraphics[width=\columnwidth]{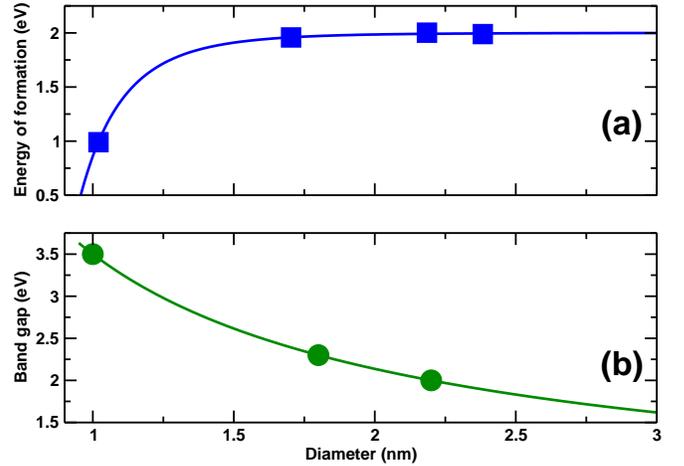}
\caption{The relative energy of formation for a vacancy in different sized 
nanocrystals is plotted in frame (a). 
In frame (b) band gap is plotted versus the crystal size.  
The solid lines show EQ.~\ref{fiteq} with the coefficients fitted to the data.
}
\label{csize}
\end{figure}

The energy of the fully optimized vacancy structures in the different 
sized nanocrystals is calculated.  Subtracting this 
energy from the energy of the perfect nanocrystals yields the energy of 
formation for a vacancy plus the chemical potential for Ge, {\it i.e.}, 
the energy to remove a Ge atom from the system.  The chemical 
potential is variable and depends on the local chemical environment.  
By assuming that all the nanocrystals are located in the same environment 
and have the same chemical potential it is possible to compare the 
relative energy of formation for vacancies in different sized 
nanocrystals.  It is known that as a nanocrystal's diameter 
approaches infinity the energy of formation approaches that found 
in bulk Ge.  Using this energy limit the calculated energy of formation 
versus crystal diameter is plotted in FIG.~\ref{csize}(a).  It is 
assumed that the size dependence goes as 
\begin{equation}\label{fiteq}
E\left(D\right)=\frac{\alpha}{D^{\beta}}+\gamma
\end{equation}
where $\alpha$, $\beta$, and $\gamma$ are fitting coefficients.  
In the limit that the diameter, $D$, goes to infinity, the energy 
equals $\gamma$.  The coefficients are determined to 
be $\alpha=-1.1395$ and $\beta=6.2574$.  The energy zero is shifted so 
that $\gamma=2.0$ eV, which is taken  
from the energies reported in TABLE \ref{libtable}.  
The quality of this fit appears good.  It is observed that the energy of 
formation is near the bulk value for crystals as small as 2.0 nm.  
This is surprising because quantum confinement continues to strongly 
influence the band gap for crystals with a similar size, as shown 
in FIG.~\ref{csize}(b).

\subsection{Distance from crystal center}

\begin{figure}
\includegraphics[width=0.8\columnwidth]{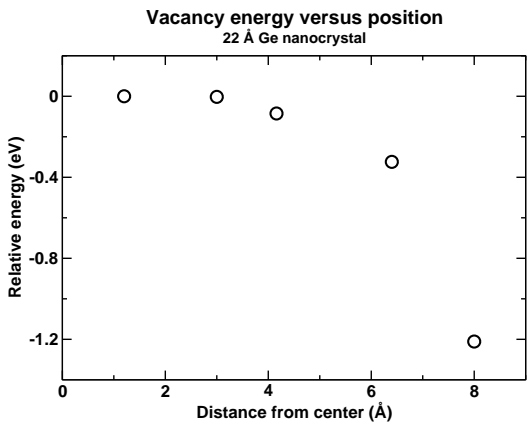}
\caption{The relative energy of a vacancy in a 2.20 nm Ge nanocrystal 
(Ge$_{244}$H$_{158}$).  The zero of the energy scale is defined to be 
the inner most atomic site.}
\label{eversusrad}
\end{figure}

To determine the influence of a vacancy's position on its energy a 2.20 nm
crystal (Ge$_{244}$H$_{158}$) is examined with a vacancy at various locations 
within it.
The calculated energies are shown in FIG.\ \ref{eversusrad}.
The configuration where the vacancy is adjacent to the center of the 
crystal [FIG.\ \ref{nanocrystals}(c)] is defined as the zero. 
Near the center of the crystal there is little change in the energy, but once
the vacancy is within 0.7 nm of the surface it begins to drop substantially.
The last stable vacancy site is 0.3 nm from the surface.
The energy of a vacancy at this site is a full 1.2 eV less than a vacancy 
near the center.
From the results in Sec.\ \ref{cryssize} it is known that the energy of
formation in the center of the 2.20 nm crystal is almost that observed in 
bulk or slightly smaller.  
Using the energies in TABLE \ref{libtable} it is deduced that the 
energy of formation for the vacancies near the surface can be no 
larger than 0.5 to 1.4 eV.  

\section{Conclusion\label{conclusion}}

A vacancy in a Ge nanocrystal undergoes a Jahn-Teller distortion.
The $T_{d}$ symmetrized broken bonds located at the vacancy introduce 
a set of three-fold degenerate, partially-occupied, states in the gap.
When these dangling bonds reconstruct the defect symmetry is lowered.  
This reduces the degeneracy of the defect states by splitting them 
into a lower-energy, fully-occupied state and two higher-energy, degenerate 
empty states.  In a bulk crystal it is known that the symmetry 
of the vacancy site is $D_{2d}$\cite{watkins1986}, but in the nanocrystals 
the surfaces introduce additional distortion.  For the smallest crystal 
the surface influence is great; the defect symmetry is $C_{s}$ and the 
energy splitting is approximately 2.7 eV.  This results in a dramatic 
reduction in energy of formation. In the 2.20 nm crystal the defect 
almost has the $D_{2d}$ symmetry that is found in bulk.  The energy 
splitting is also smaller than that in the 1.02 nm crystal, only around 
0.75 eV; therefore, the energy reduction due to the bond reconstruction 
is lower and the energy of formation is larger in the 2.20 nm crystal.  
This is consistent with the calculated prediction that the energy of 
formation will approach the bulk value for nanocrystals larger 
than 2.0 nm.  

The band gap of the crystal continues to change greatly even when 
the diameter is as large as 2 nm.  It is deduced that 
although quantum confinement continues to impact the energy levels 
in the crystal, the primary influence on the vacancy is the surface's 
ability to enhance the internal structural relaxation.  
It is concluded that in this example the quantum self-purification 
effect plays a small role if any.  A similar observation has been made 
for vacancies in Si.~\cite{beckman2007} It is hypothesized that 
this is due to the defect's wave function being highly localized 
at the reconstruction bonds.  

Finally, it is determined that vacancies placed within 0.7 nm 
of the surface are spontaneously removed.  Surprisingly vacancies 
in the interior of the crystal are stable and do not appear to be 
drawn toward the exterior.  An additional consequence is that if 
a surface were to act as a vacancy source, the vacancies produced from the 
surface are unlikely to penetrate deeply into the nanocrystal.  
The system studied here has H-passivated surfaces, which allows for 
large relaxations.  Experimental crystals that have surface 
reconstructions or polymer coatings will have more rigidity and the 
influence of the surfaces will be further muted inside the crystal.

The picture that emerges from this work is that moderate sized crystals 
will have an interior where vacancies behave bulk-like and a thin 
exterior surface region 
where the surface effects will dominate.  
Assuming that the properties of the self-interstitial defect are not strongly
modified by the surfaces, then the evidence in this paper predicts that the
self-diffusion in the interior of Ge nanocrystals will not be substantially
different from that observed in bulk.
However, recent experiments indicate that Ge surfaces are not sinks for
interstitial atoms, but instead reflect the interstitial Ge back into the
crystal.\cite{BrachtPRL2009}
If this observation holds within the nano-regime then it is possible
that the large surface to volume ratio in nanostructures will magnify
the impact of the interstitial-assisted diffusion.

\section{Acknowledgments}

The authors gratefully acknowledge support from the National Science
Foundation DMR-0755231.

\bibliography{genanobib}

\begin{thebibliography}{59}
\expandafter\ifx\csname natexlab\endcsname\relax\def\natexlab#1{#1}\fi
\expandafter\ifx\csname bibnamefont\endcsname\relax
  \def\bibnamefont#1{#1}\fi
\expandafter\ifx\csname bibfnamefont\endcsname\relax
  \def\bibfnamefont#1{#1}\fi
\expandafter\ifx\csname citenamefont\endcsname\relax
  \def\citenamefont#1{#1}\fi
\expandafter\ifx\csname url\endcsname\relax
  \def\url#1{\texttt{#1}}\fi
\expandafter\ifx\csname urlprefix\endcsname\relax\def\urlprefix{URL }\fi
\providecommand{\bibinfo}[2]{#2}
\providecommand{\eprint}[2][]{\url{#2}}

\bibitem[{\citenamefont{Yu and Cardona}(2001)}]{yu2001}
\bibinfo{author}{\bibfnamefont{P.~Y.} \bibnamefont{Yu}} \bibnamefont{and}
  \bibinfo{author}{\bibfnamefont{M.}~\bibnamefont{Cardona}},
  \emph{\bibinfo{title}{Fundamentals of Semiconductors}}
  (\bibinfo{publisher}{Springer}, \bibinfo{address}{New York},
  \bibinfo{year}{2001}), \bibinfo{edition}{3rd} ed.

\bibitem[{\citenamefont{Chui et~al.}(2003)\citenamefont{Chui, Gopalakrishnan,
  Griffin, Plummer, and Saraswat}}]{chui2003}
\bibinfo{author}{\bibfnamefont{C.~O.} \bibnamefont{Chui}},
  \bibinfo{author}{\bibfnamefont{K.}~\bibnamefont{Gopalakrishnan}},
  \bibinfo{author}{\bibfnamefont{P.~B.} \bibnamefont{Griffin}},
  \bibinfo{author}{\bibfnamefont{J.~D.} \bibnamefont{Plummer}},
  \bibnamefont{and} \bibinfo{author}{\bibfnamefont{K.~C.}
  \bibnamefont{Saraswat}}, \bibinfo{journal}{Appl. Phys. Lett.}
  \textbf{\bibinfo{volume}{83}}, \bibinfo{pages}{3275} (\bibinfo{year}{2003}).

\bibitem[{\citenamefont{Lee et~al.}(2001)\citenamefont{Lee, Leitz, Cheng,
  Pitera, Langdo, Currie, Taraschi, Fitzgerald, and Antoniadis}}]{lee2001}
\bibinfo{author}{\bibfnamefont{M.~L.} \bibnamefont{Lee}},
  \bibinfo{author}{\bibfnamefont{C.~W.} \bibnamefont{Leitz}},
  \bibinfo{author}{\bibfnamefont{Z.}~\bibnamefont{Cheng}},
  \bibinfo{author}{\bibfnamefont{A.~J.} \bibnamefont{Pitera}},
  \bibinfo{author}{\bibfnamefont{T.}~\bibnamefont{Langdo}},
  \bibinfo{author}{\bibfnamefont{M.~T.} \bibnamefont{Currie}},
  \bibinfo{author}{\bibfnamefont{G.}~\bibnamefont{Taraschi}},
  \bibinfo{author}{\bibfnamefont{E.~A.} \bibnamefont{Fitzgerald}},
  \bibnamefont{and} \bibinfo{author}{\bibfnamefont{D.~A.}
  \bibnamefont{Antoniadis}}, \bibinfo{journal}{Appl. Phys. Lett.}
  \textbf{\bibinfo{volume}{79}}, \bibinfo{pages}{3344} (\bibinfo{year}{2001}).

\bibitem[{\citenamefont{Shang et~al.}(2002)\citenamefont{Shang, Okorn-Schmidt,
  Chan, Copel, Ott, Kozlowski, Steen, Cordes, Wong, Jones et~al.}}]{shang2002}
\bibinfo{author}{\bibfnamefont{H.~L.} \bibnamefont{Shang}},
  \bibinfo{author}{\bibfnamefont{H.}~\bibnamefont{Okorn-Schmidt}},
  \bibinfo{author}{\bibfnamefont{K.~K.} \bibnamefont{Chan}},
  \bibinfo{author}{\bibfnamefont{M.}~\bibnamefont{Copel}},
  \bibinfo{author}{\bibfnamefont{J.~A.} \bibnamefont{Ott}},
  \bibinfo{author}{\bibfnamefont{P.~M.} \bibnamefont{Kozlowski}},
  \bibinfo{author}{\bibfnamefont{S.~E.} \bibnamefont{Steen}},
  \bibinfo{author}{\bibfnamefont{S.~A.} \bibnamefont{Cordes}},
  \bibinfo{author}{\bibfnamefont{H.~S.~P.} \bibnamefont{Wong}},
  \bibinfo{author}{\bibfnamefont{E.~C.} \bibnamefont{Jones}},
  \bibnamefont{et~al.}, \bibinfo{journal}{International Electron Devices 2002
  Meeting, Technical Digest} pp. \bibinfo{pages}{441--444}
  (\bibinfo{year}{2002}), \bibinfo{note}{iEEE International Electron Devices
  Meeting DEC 08-11, 2002 SAN FRANCISCO, CA IEEE Elect Devices Soc
  0-7803-7462-2}.

\bibitem[{\citenamefont{Chui et~al.}(2002)\citenamefont{Chui, Ramanathan,
  Triplett, McIntyre, and Saraswat}}]{chui2002}
\bibinfo{author}{\bibfnamefont{C.~O.} \bibnamefont{Chui}},
  \bibinfo{author}{\bibfnamefont{S.}~\bibnamefont{Ramanathan}},
  \bibinfo{author}{\bibfnamefont{B.~B.} \bibnamefont{Triplett}},
  \bibinfo{author}{\bibfnamefont{P.~C.} \bibnamefont{McIntyre}},
  \bibnamefont{and} \bibinfo{author}{\bibfnamefont{K.~C.}
  \bibnamefont{Saraswat}}, \bibinfo{journal}{IEEE Electron Device Lett.}
  \textbf{\bibinfo{volume}{23}}, \bibinfo{pages}{473} (\bibinfo{year}{2002}).

\bibitem[{\citenamefont{Lee et~al.}(2005)\citenamefont{Lee, Fitzgerald,
  Bulsara, Currie, and Lochtefeld}}]{lee2005}
\bibinfo{author}{\bibfnamefont{M.~L.} \bibnamefont{Lee}},
  \bibinfo{author}{\bibfnamefont{E.~A.} \bibnamefont{Fitzgerald}},
  \bibinfo{author}{\bibfnamefont{M.~T.} \bibnamefont{Bulsara}},
  \bibinfo{author}{\bibfnamefont{M.~T.} \bibnamefont{Currie}},
  \bibnamefont{and}
  \bibinfo{author}{\bibfnamefont{A.}~\bibnamefont{Lochtefeld}},
  \bibinfo{journal}{J. Appl. Phys.} \textbf{\bibinfo{volume}{97}}
  (\bibinfo{year}{2005}), \bibinfo{note}{011101}.

\bibitem[{\citenamefont{Shang et~al.}(2004)\citenamefont{Shang, Chu, Wang,
  Mooney, Lee, Ott, Rim, Chan, Guarini, and Ieong}}]{shang2004}
\bibinfo{author}{\bibfnamefont{H.}~\bibnamefont{Shang}},
  \bibinfo{author}{\bibfnamefont{J.~O.} \bibnamefont{Chu}},
  \bibinfo{author}{\bibfnamefont{X.}~\bibnamefont{Wang}},
  \bibinfo{author}{\bibfnamefont{P.~M.} \bibnamefont{Mooney}},
  \bibinfo{author}{\bibfnamefont{K.}~\bibnamefont{Lee}},
  \bibinfo{author}{\bibfnamefont{J.}~\bibnamefont{Ott}},
  \bibinfo{author}{\bibfnamefont{K.}~\bibnamefont{Rim}},
  \bibinfo{author}{\bibfnamefont{K.}~\bibnamefont{Chan}},
  \bibinfo{author}{\bibfnamefont{K.}~\bibnamefont{Guarini}}, \bibnamefont{and}
  \bibinfo{author}{\bibfnamefont{M.}~\bibnamefont{Ieong}},
  \bibinfo{journal}{2004 Symposium on Vlsi Technology, Digest of Technical
  Papers} pp. \bibinfo{pages}{204--205} (\bibinfo{year}{2004}),
  \bibinfo{note}{symposium on VLSI Technology JUN 15-17, 2004 Honolulu, HI IEEE
  Electron Devices Soc, Japan Soc Appl Phys 0-7803-8289-7}.

\bibitem[{\citenamefont{Shang et~al.}(2006)\citenamefont{Shang, Frank, Gusev,
  Chu, Bedell, Guarini, and Ieong}}]{shang2006}
\bibinfo{author}{\bibfnamefont{H.}~\bibnamefont{Shang}},
  \bibinfo{author}{\bibfnamefont{M.~M.} \bibnamefont{Frank}},
  \bibinfo{author}{\bibfnamefont{E.~P.} \bibnamefont{Gusev}},
  \bibinfo{author}{\bibfnamefont{J.~O.} \bibnamefont{Chu}},
  \bibinfo{author}{\bibfnamefont{S.~W.} \bibnamefont{Bedell}},
  \bibinfo{author}{\bibfnamefont{K.~W.} \bibnamefont{Guarini}},
  \bibnamefont{and} \bibinfo{author}{\bibfnamefont{M.}~\bibnamefont{Ieong}},
  \bibinfo{journal}{IBM J. Res. and Dev.} \textbf{\bibinfo{volume}{50}},
  \bibinfo{pages}{377} (\bibinfo{year}{2006}).

\bibitem[{\citenamefont{Cheng et~al.}(2004)\citenamefont{Cheng, Jung, Lee,
  Pitera, Hoyt, Antoniadis, and Fitzgerald}}]{cheng2004}
\bibinfo{author}{\bibfnamefont{Z.~Y.} \bibnamefont{Cheng}},
  \bibinfo{author}{\bibfnamefont{J.~W.} \bibnamefont{Jung}},
  \bibinfo{author}{\bibfnamefont{M.~L.} \bibnamefont{Lee}},
  \bibinfo{author}{\bibfnamefont{A.~J.} \bibnamefont{Pitera}},
  \bibinfo{author}{\bibfnamefont{J.~L.} \bibnamefont{Hoyt}},
  \bibinfo{author}{\bibfnamefont{D.~A.} \bibnamefont{Antoniadis}},
  \bibnamefont{and} \bibinfo{author}{\bibfnamefont{E.~A.}
  \bibnamefont{Fitzgerald}}, \bibinfo{journal}{Semicond. Sci. Technol.}
  \textbf{\bibinfo{volume}{19}}, \bibinfo{pages}{L48} (\bibinfo{year}{2004}).

\bibitem[{\citenamefont{Queisser and Haller}(1998)}]{queisser1998}
\bibinfo{author}{\bibfnamefont{H.~J.} \bibnamefont{Queisser}} \bibnamefont{and}
  \bibinfo{author}{\bibfnamefont{E.~E.} \bibnamefont{Haller}},
  \bibinfo{journal}{Science} \textbf{\bibinfo{volume}{281}},
  \bibinfo{pages}{945} (\bibinfo{year}{1998}).

\bibitem[{\citenamefont{Bracht}(2000)}]{bracht2000}
\bibinfo{author}{\bibfnamefont{H.}~\bibnamefont{Bracht}}, \bibinfo{journal}{Mrs
  Bulletin} \textbf{\bibinfo{volume}{25}}, \bibinfo{pages}{22}
  (\bibinfo{year}{2000}).

\bibitem[{\citenamefont{Ganster et~al.}(2009)\citenamefont{Ganster, Treglia,
  and Saul}}]{ganster2009}
\bibinfo{author}{\bibfnamefont{P.}~\bibnamefont{Ganster}},
  \bibinfo{author}{\bibfnamefont{G.}~\bibnamefont{Treglia}}, \bibnamefont{and}
  \bibinfo{author}{\bibfnamefont{A.}~\bibnamefont{Saul}},
  \bibinfo{journal}{Phys. Rev. B} \textbf{\bibinfo{volume}{79}}
  (\bibinfo{year}{2009}), \bibinfo{note}{115205}.

\bibitem[{\citenamefont{Shimizu et~al.}(2007)\citenamefont{Shimizu, Uematsu,
  and Itoh}}]{shimizu2007}
\bibinfo{author}{\bibfnamefont{Y.}~\bibnamefont{Shimizu}},
  \bibinfo{author}{\bibfnamefont{M.}~\bibnamefont{Uematsu}}, \bibnamefont{and}
  \bibinfo{author}{\bibfnamefont{K.~M.} \bibnamefont{Itoh}},
  \bibinfo{journal}{Phys. Rev. Lett.} \textbf{\bibinfo{volume}{98}},
  \bibinfo{pages}{095901} (\bibinfo{year}{2007}), \bibinfo{note}{and references
  therein}.

\bibitem[{\citenamefont{Silvestri et~al.}(2006)\citenamefont{Silvestri, Bracht,
  Hansen, Larsen, and Haller}}]{silvestry2006}
\bibinfo{author}{\bibfnamefont{H.~H.} \bibnamefont{Silvestri}},
  \bibinfo{author}{\bibfnamefont{H.}~\bibnamefont{Bracht}},
  \bibinfo{author}{\bibfnamefont{J.~L.} \bibnamefont{Hansen}},
  \bibinfo{author}{\bibfnamefont{A.~N.} \bibnamefont{Larsen}},
  \bibnamefont{and} \bibinfo{author}{\bibfnamefont{E.~E.}
  \bibnamefont{Haller}}, \bibinfo{journal}{Semicond. Sci. Technol.}
  \textbf{\bibinfo{volume}{21}}, \bibinfo{pages}{758} (\bibinfo{year}{2006}).

\bibitem[{\citenamefont{Bracht et~al.}(2009)\citenamefont{Bracht, Schneider,
  Klug, Liao, Hansen, Haller, Larsen, Bougeard, Posselt, and
  W\"undisch}}]{BrachtPRL2009}
\bibinfo{author}{\bibfnamefont{H.}~\bibnamefont{Bracht}},
  \bibinfo{author}{\bibfnamefont{S.}~\bibnamefont{Schneider}},
  \bibinfo{author}{\bibfnamefont{J.~N.} \bibnamefont{Klug}},
  \bibinfo{author}{\bibfnamefont{C.~Y.} \bibnamefont{Liao}},
  \bibinfo{author}{\bibfnamefont{J.~L.} \bibnamefont{Hansen}},
  \bibinfo{author}{\bibfnamefont{E.~E.} \bibnamefont{Haller}},
  \bibinfo{author}{\bibfnamefont{A.~N.} \bibnamefont{Larsen}},
  \bibinfo{author}{\bibfnamefont{D.}~\bibnamefont{Bougeard}},
  \bibinfo{author}{\bibfnamefont{M.}~\bibnamefont{Posselt}}, \bibnamefont{and}
  \bibinfo{author}{\bibfnamefont{C.}~\bibnamefont{W\"undisch}},
  \bibinfo{journal}{Phys. Rev. Lett.} \textbf{\bibinfo{volume}{103}},
  \bibinfo{pages}{255501} (\bibinfo{year}{2009}).

\bibitem[{\citenamefont{Zhu et~al.}(1996)\citenamefont{Zhu, de~la Rubia, Yang,
  Mailhiot, and Gilmer}}]{zhu1996}
\bibinfo{author}{\bibfnamefont{J.}~\bibnamefont{Zhu}},
  \bibinfo{author}{\bibfnamefont{T.~D.} \bibnamefont{de~la Rubia}},
  \bibinfo{author}{\bibfnamefont{L.~H.} \bibnamefont{Yang}},
  \bibinfo{author}{\bibfnamefont{C.}~\bibnamefont{Mailhiot}}, \bibnamefont{and}
  \bibinfo{author}{\bibfnamefont{G.~H.} \bibnamefont{Gilmer}},
  \bibinfo{journal}{Phys. Rev. B} \textbf{\bibinfo{volume}{54}},
  \bibinfo{pages}{4741} (\bibinfo{year}{1996}).

\bibitem[{\citenamefont{Sahli and Fichtner}(2005)}]{sahli2005}
\bibinfo{author}{\bibfnamefont{B.}~\bibnamefont{Sahli}} \bibnamefont{and}
  \bibinfo{author}{\bibfnamefont{W.}~\bibnamefont{Fichtner}},
  \bibinfo{journal}{Phys. Rev. B} \textbf{\bibinfo{volume}{72}}
  (\bibinfo{year}{2005}).

\bibitem[{\citenamefont{Centoni et~al.}(2005)\citenamefont{Centoni, Sadigh,
  Gilmer, Lenosky, de~la Rubia, and Musgrave}}]{centoni2005}
\bibinfo{author}{\bibfnamefont{S.~A.} \bibnamefont{Centoni}},
  \bibinfo{author}{\bibfnamefont{B.}~\bibnamefont{Sadigh}},
  \bibinfo{author}{\bibfnamefont{G.~H.} \bibnamefont{Gilmer}},
  \bibinfo{author}{\bibfnamefont{T.~J.} \bibnamefont{Lenosky}},
  \bibinfo{author}{\bibfnamefont{T.~D.} \bibnamefont{de~la Rubia}},
  \bibnamefont{and} \bibinfo{author}{\bibfnamefont{C.~B.}
  \bibnamefont{Musgrave}}, \bibinfo{journal}{Phys. Rev. B}
  \textbf{\bibinfo{volume}{72}} (\bibinfo{year}{2005}).

\bibitem[{\citenamefont{Leung et~al.}(1999)\citenamefont{Leung, Needs,
  Rajagopal, Itoh, and Ihara}}]{leung1999}
\bibinfo{author}{\bibfnamefont{W.~K.} \bibnamefont{Leung}},
  \bibinfo{author}{\bibfnamefont{R.~J.} \bibnamefont{Needs}},
  \bibinfo{author}{\bibfnamefont{G.}~\bibnamefont{Rajagopal}},
  \bibinfo{author}{\bibfnamefont{S.}~\bibnamefont{Itoh}}, \bibnamefont{and}
  \bibinfo{author}{\bibfnamefont{S.}~\bibnamefont{Ihara}},
  \bibinfo{journal}{Phys. Rev. Lett.} \textbf{\bibinfo{volume}{83}},
  \bibinfo{pages}{2351} (\bibinfo{year}{1999}).

\bibitem[{\citenamefont{Antonelli et~al.}(1998)\citenamefont{Antonelli,
  Kaxiras, and Chadi}}]{antonelli1998}
\bibinfo{author}{\bibfnamefont{A.}~\bibnamefont{Antonelli}},
  \bibinfo{author}{\bibfnamefont{E.}~\bibnamefont{Kaxiras}}, \bibnamefont{and}
  \bibinfo{author}{\bibfnamefont{D.~J.} \bibnamefont{Chadi}},
  \bibinfo{journal}{Phys. Rev. Lett.} \textbf{\bibinfo{volume}{81}},
  \bibinfo{pages}{2088} (\bibinfo{year}{1998}).

\bibitem[{\citenamefont{Wang et~al.}(1991)\citenamefont{Wang, Chan, and
  Ho}}]{wang1991}
\bibinfo{author}{\bibfnamefont{C.~Z.} \bibnamefont{Wang}},
  \bibinfo{author}{\bibfnamefont{C.~T.} \bibnamefont{Chan}}, \bibnamefont{and}
  \bibinfo{author}{\bibfnamefont{K.~M.} \bibnamefont{Ho}},
  \bibinfo{journal}{Phys. Rev. Lett.} \textbf{\bibinfo{volume}{66}},
  \bibinfo{pages}{189} (\bibinfo{year}{1991}).

\bibitem[{\citenamefont{Wright}(2006)}]{wright2006}
\bibinfo{author}{\bibfnamefont{A.~F.} \bibnamefont{Wright}},
  \bibinfo{journal}{Phys. Rev. B} \textbf{\bibinfo{volume}{74}}
  (\bibinfo{year}{2006}).

\bibitem[{\citenamefont{Fazzio et~al.}(2000)\citenamefont{Fazzio, Janotti,
  da~Silva, and Mota}}]{fazzio2000}
\bibinfo{author}{\bibfnamefont{A.}~\bibnamefont{Fazzio}},
  \bibinfo{author}{\bibfnamefont{A.}~\bibnamefont{Janotti}},
  \bibinfo{author}{\bibfnamefont{A.~J.~R.} \bibnamefont{da~Silva}},
  \bibnamefont{and} \bibinfo{author}{\bibfnamefont{R.}~\bibnamefont{Mota}},
  \bibinfo{journal}{Phys. Rev. B} \textbf{\bibinfo{volume}{61}},
  \bibinfo{pages}{R2401} (\bibinfo{year}{2000}).

\bibitem[{\citenamefont{da~Silva et~al.}(2000)\citenamefont{da~Silva, Janotti,
  Fazzio, Baierle, and Mota}}]{dasilva2000}
\bibinfo{author}{\bibfnamefont{A.~J.~R.} \bibnamefont{da~Silva}},
  \bibinfo{author}{\bibfnamefont{A.}~\bibnamefont{Janotti}},
  \bibinfo{author}{\bibfnamefont{A.}~\bibnamefont{Fazzio}},
  \bibinfo{author}{\bibfnamefont{R.~J.} \bibnamefont{Baierle}},
  \bibnamefont{and} \bibinfo{author}{\bibfnamefont{R.}~\bibnamefont{Mota}},
  \bibinfo{journal}{Phys. Rev. B} \textbf{\bibinfo{volume}{62}},
  \bibinfo{pages}{9903} (\bibinfo{year}{2000}).

\bibitem[{\citenamefont{Uberuaga et~al.}(2002)\citenamefont{Uberuaga,
  Henkelman, Jonsson, Dunham, Windl, and Stumpf}}]{uberuaga2002}
\bibinfo{author}{\bibfnamefont{B.~P.} \bibnamefont{Uberuaga}},
  \bibinfo{author}{\bibfnamefont{G.}~\bibnamefont{Henkelman}},
  \bibinfo{author}{\bibfnamefont{H.}~\bibnamefont{Jonsson}},
  \bibinfo{author}{\bibfnamefont{S.~T.} \bibnamefont{Dunham}},
  \bibinfo{author}{\bibfnamefont{W.}~\bibnamefont{Windl}}, \bibnamefont{and}
  \bibinfo{author}{\bibfnamefont{R.}~\bibnamefont{Stumpf}},
  \bibinfo{journal}{Physica Status Solidi B-Basic Research}
  \textbf{\bibinfo{volume}{233}}, \bibinfo{pages}{24} (\bibinfo{year}{2002}),
  \bibinfo{note}{3rd Motorola Workshop on Computational Materials and
  Electronics NOV 14-16, 2001 TEMPE, ARIZONA}.

\bibitem[{\citenamefont{Carvalho et~al.}(2007)\citenamefont{Carvalho, Jones,
  Janke, Goss, Briddon, Coutinho, and \"Oberg}}]{CarvalhoPRL2007}
\bibinfo{author}{\bibfnamefont{A.}~\bibnamefont{Carvalho}},
  \bibinfo{author}{\bibfnamefont{R.}~\bibnamefont{Jones}},
  \bibinfo{author}{\bibfnamefont{C.}~\bibnamefont{Janke}},
  \bibinfo{author}{\bibfnamefont{J.~P.} \bibnamefont{Goss}},
  \bibinfo{author}{\bibfnamefont{P.~R.} \bibnamefont{Briddon}},
  \bibinfo{author}{\bibfnamefont{J.}~\bibnamefont{Coutinho}}, \bibnamefont{and}
  \bibinfo{author}{\bibfnamefont{S.}~\bibnamefont{\"Oberg}},
  \bibinfo{journal}{Phys. Rev. Lett.} \textbf{\bibinfo{volume}{99}},
  \bibinfo{pages}{175502} (\bibinfo{year}{2007}).

\bibitem[{\citenamefont{Hwang and Watt}(1968)}]{hwang1968}
\bibinfo{author}{\bibfnamefont{C.~J.} \bibnamefont{Hwang}} \bibnamefont{and}
  \bibinfo{author}{\bibfnamefont{L.~A.~K.} \bibnamefont{Watt}},
  \bibinfo{journal}{Phys. Rev.} \textbf{\bibinfo{volume}{171}},
  \bibinfo{pages}{958} (\bibinfo{year}{1968}).

\bibitem[{\citenamefont{Chroneos
  et~al.}(2008{\natexlab{a}})\citenamefont{Chroneos, Grimes, Uberuaga, and
  Bracht}}]{chroneos2008}
\bibinfo{author}{\bibfnamefont{A.}~\bibnamefont{Chroneos}},
  \bibinfo{author}{\bibfnamefont{R.~W.} \bibnamefont{Grimes}},
  \bibinfo{author}{\bibfnamefont{B.~P.} \bibnamefont{Uberuaga}},
  \bibnamefont{and} \bibinfo{author}{\bibfnamefont{H.}~\bibnamefont{Bracht}},
  \bibinfo{journal}{Phys. Rev. B} \textbf{\bibinfo{volume}{77}},
  \bibinfo{pages}{235208} (\bibinfo{year}{2008}{\natexlab{a}}),
  \bibinfo{note}{and references therein}.

\bibitem[{\citenamefont{Pinto et~al.}(2006)\citenamefont{Pinto, Coutinho,
  Torres, Öberg, and Briddon}}]{Pinto2006498}
\bibinfo{author}{\bibfnamefont{H.}~\bibnamefont{Pinto}},
  \bibinfo{author}{\bibfnamefont{J.}~\bibnamefont{Coutinho}},
  \bibinfo{author}{\bibfnamefont{V.}~\bibnamefont{Torres}},
  \bibinfo{author}{\bibfnamefont{S.}~\bibnamefont{Öberg}}, \bibnamefont{and}
  \bibinfo{author}{\bibfnamefont{P.}~\bibnamefont{Briddon}},
  \bibinfo{journal}{Materials Science in Semiconductor Processing}
  \textbf{\bibinfo{volume}{9}}, \bibinfo{pages}{498 } (\bibinfo{year}{2006}),
  ISSN \bibinfo{issn}{1369-8001}, \bibinfo{note}{proceedings of Symposium T
  E-MRS 2006 Spring Meeting on Germanium based semiconductors from materials to
  devices},
  \urlprefix\url{http://www.sciencedirect.com/science/article/B6VPK-4M340JT-4/%
2/2ceedc81773b2a226212cb616b9e0c1e}.

\bibitem[{\citenamefont{Moreira et~al.}(2004)\citenamefont{Moreira, Miwa, and
  Venezuela}}]{Moreira2004}
\bibinfo{author}{\bibfnamefont{M.~D.} \bibnamefont{Moreira}},
  \bibinfo{author}{\bibfnamefont{R.~H.} \bibnamefont{Miwa}}, \bibnamefont{and}
  \bibinfo{author}{\bibfnamefont{P.}~\bibnamefont{Venezuela}},
  \bibinfo{journal}{Phys. Rev. B} \textbf{\bibinfo{volume}{70}},
  \bibinfo{pages}{115215} (\bibinfo{year}{2004}).

\bibitem[{\citenamefont{Vanhellemont et~al.}(2007)\citenamefont{Vanhellemont,
  \'{S}piewak, and Sueoka}}]{Vanhellemont2007}
\bibinfo{author}{\bibfnamefont{J.}~\bibnamefont{Vanhellemont}},
  \bibinfo{author}{\bibfnamefont{P.}~\bibnamefont{\'{S}piewak}},
  \bibnamefont{and} \bibinfo{author}{\bibfnamefont{K.}~\bibnamefont{Sueoka}},
  \bibinfo{journal}{Journal of Applied Physics} \textbf{\bibinfo{volume}{101}},
  \bibinfo{eid}{036103} (pages~\bibinfo{numpages}{3}) (\bibinfo{year}{2007}),
  \urlprefix\url{http://link.aip.org/link/?JAP/101/036103/1}.

\bibitem[{\citenamefont{Werner et~al.}(1985)\citenamefont{Werner, Mehrer, and
  Hochheimer}}]{werner1985}
\bibinfo{author}{\bibfnamefont{M.}~\bibnamefont{Werner}},
  \bibinfo{author}{\bibfnamefont{H.}~\bibnamefont{Mehrer}}, \bibnamefont{and}
  \bibinfo{author}{\bibfnamefont{H.~D.} \bibnamefont{Hochheimer}},
  \bibinfo{journal}{Phys. Rev. B} \textbf{\bibinfo{volume}{32}},
  \bibinfo{pages}{3930} (\bibinfo{year}{1985}).

\bibitem[{\citenamefont{Bracht et~al.}(2007)\citenamefont{Bracht, Silvestri,
  Sharp, and Haller}}]{bracht2007}
\bibinfo{author}{\bibfnamefont{H.}~\bibnamefont{Bracht}},
  \bibinfo{author}{\bibfnamefont{H.~H.} \bibnamefont{Silvestri}},
  \bibinfo{author}{\bibfnamefont{I.~D.} \bibnamefont{Sharp}}, \bibnamefont{and}
  \bibinfo{author}{\bibfnamefont{E.~E.} \bibnamefont{Haller}},
  \bibinfo{journal}{Phys. Rev. B} \textbf{\bibinfo{volume}{75}}
  (\bibinfo{year}{2007}), \bibinfo{note}{035211}.

\bibitem[{\citenamefont{Bracht et~al.}(1991)\citenamefont{Bracht, Stolwijk, and
  Mehrer}}]{bracht1991}
\bibinfo{author}{\bibfnamefont{H.}~\bibnamefont{Bracht}},
  \bibinfo{author}{\bibfnamefont{N.~A.} \bibnamefont{Stolwijk}},
  \bibnamefont{and} \bibinfo{author}{\bibfnamefont{H.}~\bibnamefont{Mehrer}},
  \bibinfo{journal}{Phys. Rev. B} \textbf{\bibinfo{volume}{43}},
  \bibinfo{pages}{14465} (\bibinfo{year}{1991}).

\bibitem[{\citenamefont{Brotzmann et~al.}(2008)\citenamefont{Brotzmann, Bracht,
  Hansen, Larsen, Simoen, Haller, and Christensen}}]{brotzmann2008}
\bibinfo{author}{\bibfnamefont{S.}~\bibnamefont{Brotzmann}},
  \bibinfo{author}{\bibfnamefont{H.}~\bibnamefont{Bracht}},
  \bibinfo{author}{\bibfnamefont{J.~L.} \bibnamefont{Hansen}},
  \bibinfo{author}{\bibfnamefont{A.~N.} \bibnamefont{Larsen}},
  \bibinfo{author}{\bibfnamefont{E.}~\bibnamefont{Simoen}},
  \bibinfo{author}{\bibfnamefont{E.~E.} \bibnamefont{Haller}},
  \bibnamefont{and}
  \bibinfo{author}{\bibfnamefont{P.}~\bibnamefont{Christensen},
  \bibfnamefont{J.~S. an d~Werner}}, \bibinfo{journal}{Phys. Rev. B}
  \textbf{\bibinfo{volume}{77}} (\bibinfo{year}{2008}), \bibinfo{note}{235207}.

\bibitem[{\citenamefont{Chroneos
  et~al.}(2008{\natexlab{b}})\citenamefont{Chroneos, Bracht, Grimes, and
  Uberuaga}}]{chroneos20082}
\bibinfo{author}{\bibfnamefont{A.}~\bibnamefont{Chroneos}},
  \bibinfo{author}{\bibfnamefont{H.}~\bibnamefont{Bracht}},
  \bibinfo{author}{\bibfnamefont{R.~W.} \bibnamefont{Grimes}},
  \bibnamefont{and} \bibinfo{author}{\bibfnamefont{B.~P.}
  \bibnamefont{Uberuaga}}, \bibinfo{journal}{Appl. Phys. Lett.}
  \textbf{\bibinfo{volume}{92}} (\bibinfo{year}{2008}{\natexlab{b}}),
  \bibinfo{note}{172103}.

\bibitem[{\citenamefont{Das et~al.}(2009)\citenamefont{Das, Singha, Das, Dhar,
  and Ray}}]{das2009}
\bibinfo{author}{\bibfnamefont{S.}~\bibnamefont{Das}},
  \bibinfo{author}{\bibfnamefont{R.~K.} \bibnamefont{Singha}},
  \bibinfo{author}{\bibfnamefont{K.}~\bibnamefont{Das}},
  \bibinfo{author}{\bibfnamefont{A.}~\bibnamefont{Dhar}}, \bibnamefont{and}
  \bibinfo{author}{\bibfnamefont{S.~K.} \bibnamefont{Ray}},
  \bibinfo{journal}{Journal of Nanoscience and Nanotechnology}
  \textbf{\bibinfo{volume}{9}}, \bibinfo{pages}{5484} (\bibinfo{year}{2009}),
  \bibinfo{note}{international Conference on Nanoscience and Technology FEB
  27-29, 2008 Chennai, INDIA}.

\bibitem[{\citenamefont{Ma and Wang}(2008)}]{ma2008}
\bibinfo{author}{\bibfnamefont{X.}~\bibnamefont{Ma}} \bibnamefont{and}
  \bibinfo{author}{\bibfnamefont{C.}~\bibnamefont{Wang}},
  \bibinfo{journal}{Appl. Phys. B} \textbf{\bibinfo{volume}{92}},
  \bibinfo{pages}{589} (\bibinfo{year}{2008}).

\bibitem[{\citenamefont{Sato et~al.}(1995)\citenamefont{Sato, Morisaki, and
  Iwase}}]{sato1995}
\bibinfo{author}{\bibfnamefont{S.}~\bibnamefont{Sato}},
  \bibinfo{author}{\bibfnamefont{H.}~\bibnamefont{Morisaki}}, \bibnamefont{and}
  \bibinfo{author}{\bibfnamefont{M.}~\bibnamefont{Iwase}},
  \bibinfo{journal}{Appl. Phys. Lett.} \textbf{\bibinfo{volume}{66}},
  \bibinfo{pages}{3176} (\bibinfo{year}{1995}).

\bibitem[{\citenamefont{Zhang et~al.}(2008)\citenamefont{Zhang, He, Zhang, Liu,
  Fu, Song, and Zhang}}]{zhang2008}
\bibinfo{author}{\bibfnamefont{L.~N.} \bibnamefont{Zhang}},
  \bibinfo{author}{\bibfnamefont{J.}~\bibnamefont{He}},
  \bibinfo{author}{\bibfnamefont{J.}~\bibnamefont{Zhang}},
  \bibinfo{author}{\bibfnamefont{F.}~\bibnamefont{Liu}},
  \bibinfo{author}{\bibfnamefont{Y.}~\bibnamefont{Fu}},
  \bibinfo{author}{\bibfnamefont{Y.}~\bibnamefont{Song}}, \bibnamefont{and}
  \bibinfo{author}{\bibfnamefont{X.}~\bibnamefont{Zhang}},
  \bibinfo{journal}{IEEE Trans. Electron Devices}
  \textbf{\bibinfo{volume}{55}}, \bibinfo{pages}{2907} (\bibinfo{year}{2008}).

\bibitem[{\citenamefont{Jiang et~al.}(2008)\citenamefont{Jiang, Singh, Liow,
  Loh, Balakumar, Hoe, Tung, Bliznetsov, Rustagi, Lo et~al.}}]{jiang2008}
\bibinfo{author}{\bibfnamefont{Y.}~\bibnamefont{Jiang}},
  \bibinfo{author}{\bibfnamefont{N.}~\bibnamefont{Singh}},
  \bibinfo{author}{\bibfnamefont{T.~Y.} \bibnamefont{Liow}},
  \bibinfo{author}{\bibfnamefont{W.~Y.} \bibnamefont{Loh}},
  \bibinfo{author}{\bibfnamefont{S.}~\bibnamefont{Balakumar}},
  \bibinfo{author}{\bibfnamefont{K.~M.} \bibnamefont{Hoe}},
  \bibinfo{author}{\bibfnamefont{C.~H.} \bibnamefont{Tung}},
  \bibinfo{author}{\bibfnamefont{V.}~\bibnamefont{Bliznetsov}},
  \bibinfo{author}{\bibfnamefont{S.~C.} \bibnamefont{Rustagi}},
  \bibinfo{author}{\bibfnamefont{G.~Q.} \bibnamefont{Lo}},
  \bibnamefont{et~al.}, \bibinfo{journal}{IEEE Electron Device Lett.}
  \textbf{\bibinfo{volume}{29}}, \bibinfo{pages}{595} (\bibinfo{year}{2008}).

\bibitem[{\citenamefont{Markussen et~al.}(2007)\citenamefont{Markussen, Rurali,
  Jauho, and Brandbyge}}]{markussen2007}
\bibinfo{author}{\bibfnamefont{T.}~\bibnamefont{Markussen}},
  \bibinfo{author}{\bibfnamefont{R.}~\bibnamefont{Rurali}},
  \bibinfo{author}{\bibfnamefont{A.~P.} \bibnamefont{Jauho}}, \bibnamefont{and}
  \bibinfo{author}{\bibfnamefont{M.}~\bibnamefont{Brandbyge}},
  \bibinfo{journal}{Phys. Rev. Lett.} \textbf{\bibinfo{volume}{99}}
  (\bibinfo{year}{2007}), \bibinfo{note}{076803}.

\bibitem[{\citenamefont{Beckman et~al.}(2006)\citenamefont{Beckman, Han, and
  Chelikowsky}}]{beckman2006}
\bibinfo{author}{\bibfnamefont{S.~P.} \bibnamefont{Beckman}},
  \bibinfo{author}{\bibfnamefont{J.~X.} \bibnamefont{Han}}, \bibnamefont{and}
  \bibinfo{author}{\bibfnamefont{J.~R.} \bibnamefont{Chelikowsky}},
  \bibinfo{journal}{Phys. Rev. B} \textbf{\bibinfo{volume}{74}}
  (\bibinfo{year}{2006}), \bibinfo{note}{165314}.

\bibitem[{\citenamefont{Dalpian and Chelikowsky}(2006)}]{dalpian2006}
\bibinfo{author}{\bibfnamefont{G.~M.} \bibnamefont{Dalpian}} \bibnamefont{and}
  \bibinfo{author}{\bibfnamefont{J.~R.} \bibnamefont{Chelikowsky}},
  \bibinfo{journal}{Phys. Rev. Lett.} \textbf{\bibinfo{volume}{96}}
  (\bibinfo{year}{2006}), \bibinfo{note}{226802}.

\bibitem[{\citenamefont{Du et~al.}(2008)\citenamefont{Du, Erwin, Efros, and
  Norris}}]{du2008}
\bibinfo{author}{\bibfnamefont{M.~H.} \bibnamefont{Du}},
  \bibinfo{author}{\bibfnamefont{S.~C.} \bibnamefont{Erwin}},
  \bibinfo{author}{\bibfnamefont{A.~L.} \bibnamefont{Efros}}, \bibnamefont{and}
  \bibinfo{author}{\bibfnamefont{D.~J.} \bibnamefont{Norris}},
  \bibinfo{journal}{Phys. Rev. Lett.} \textbf{\bibinfo{volume}{100}}
  (\bibinfo{year}{2008}), \bibinfo{note}{179702}.

\bibitem[{\citenamefont{Dalpian and Chelikowsky}(2008)}]{dalpian2008}
\bibinfo{author}{\bibfnamefont{G.~M.} \bibnamefont{Dalpian}} \bibnamefont{and}
  \bibinfo{author}{\bibfnamefont{J.~R.} \bibnamefont{Chelikowsky}},
  \bibinfo{journal}{Phys. Rev. Lett.} \textbf{\bibinfo{volume}{100}}
  (\bibinfo{year}{2008}), \bibinfo{note}{179703}.

\bibitem[{\citenamefont{Li et~al.}(2008)\citenamefont{Li, Wei, Li, and
  Xia}}]{li2008}
\bibinfo{author}{\bibfnamefont{J.~B.} \bibnamefont{Li}},
  \bibinfo{author}{\bibfnamefont{S.~H.} \bibnamefont{Wei}},
  \bibinfo{author}{\bibfnamefont{S.~S.} \bibnamefont{Li}}, \bibnamefont{and}
  \bibinfo{author}{\bibfnamefont{J.~B.} \bibnamefont{Xia}},
  \bibinfo{journal}{Phys. Rev. B} \textbf{\bibinfo{volume}{77}}
  (\bibinfo{year}{2008}), \bibinfo{note}{113304}.

\bibitem[{\citenamefont{Beckman and
  Chelikowsky}(2007{\natexlab{a}})}]{beckman2007}
\bibinfo{author}{\bibfnamefont{S.~P.} \bibnamefont{Beckman}} \bibnamefont{and}
  \bibinfo{author}{\bibfnamefont{J.~R.} \bibnamefont{Chelikowsky}},
  \bibinfo{journal}{Physica B} \textbf{\bibinfo{volume}{401}},
  \bibinfo{pages}{537} (\bibinfo{year}{2007}{\natexlab{a}}),
  \bibinfo{note}{24th International Conference on Defects in Semiconductors JUL
  22-27, 2007 Albuquerque, NM}.

\bibitem[{\citenamefont{Beckman and
  Chelikowsky}(2007{\natexlab{b}})}]{beckunpub}
\bibinfo{author}{\bibfnamefont{S.~P.} \bibnamefont{Beckman}} \bibnamefont{and}
  \bibinfo{author}{\bibfnamefont{J.~R.} \bibnamefont{Chelikowsky}},
  \bibinfo{journal}{unpublished data}  (\bibinfo{year}{2007}{\natexlab{b}}).

\bibitem[{\citenamefont{Kohn and Sham}(1965)}]{DFT}
\bibinfo{author}{\bibfnamefont{W.}~\bibnamefont{Kohn}} \bibnamefont{and}
  \bibinfo{author}{\bibfnamefont{L.~J.} \bibnamefont{Sham}},
  \bibinfo{journal}{Phys. Rev.} \textbf{\bibinfo{volume}{140}},
  \bibinfo{pages}{A1133} (\bibinfo{year}{1965}).

\bibitem[{\citenamefont{Ceperley and Alder}(1980)}]{LDA}
\bibinfo{author}{\bibfnamefont{D.~M.} \bibnamefont{Ceperley}} \bibnamefont{and}
  \bibinfo{author}{\bibfnamefont{B.~J.} \bibnamefont{Alder}},
  \bibinfo{journal}{Phys. Rev. Lett.} \textbf{\bibinfo{volume}{45}},
  \bibinfo{pages}{566} (\bibinfo{year}{1980}).

\bibitem[{\citenamefont{Soler et~al.}(2002)\citenamefont{Soler, Artacho, Gale,
  Garc\'{\i}a, Junquera, Ordej\'on, and S\'anchez-Portal}}]{Siesta}
\bibinfo{author}{\bibfnamefont{J.~M.} \bibnamefont{Soler}},
  \bibinfo{author}{\bibfnamefont{E.}~\bibnamefont{Artacho}},
  \bibinfo{author}{\bibfnamefont{J.~D.} \bibnamefont{Gale}},
  \bibinfo{author}{\bibfnamefont{A.}~\bibnamefont{Garc\'{\i}a}},
  \bibinfo{author}{\bibfnamefont{J.}~\bibnamefont{Junquera}},
  \bibinfo{author}{\bibfnamefont{P.}~\bibnamefont{Ordej\'on}},
  \bibnamefont{and}
  \bibinfo{author}{\bibfnamefont{D.}~\bibnamefont{S\'anchez-Portal}},
  \bibinfo{journal}{J. Phys.: Condens. Matter} \textbf{\bibinfo{volume}{14}},
  \bibinfo{pages}{2745} (\bibinfo{year}{2002}).

\bibitem[{\citenamefont{Troullier and Martins}(1991)}]{TroullierMartins}
\bibinfo{author}{\bibfnamefont{N.}~\bibnamefont{Troullier}} \bibnamefont{and}
  \bibinfo{author}{\bibfnamefont{J.~L.} \bibnamefont{Martins}},
  \bibinfo{journal}{Phys. Rev. B} \textbf{\bibinfo{volume}{43}},
  \bibinfo{pages}{1993} (\bibinfo{year}{1991}).

\bibitem[{\citenamefont{Anglada et~al.}(2002)\citenamefont{Anglada, Soler,
  Junquera, and Artacho}}]{BasisSets}
\bibinfo{author}{\bibfnamefont{E.}~\bibnamefont{Anglada}},
  \bibinfo{author}{\bibfnamefont{J.~M.} \bibnamefont{Soler}},
  \bibinfo{author}{\bibfnamefont{J.}~\bibnamefont{Junquera}}, \bibnamefont{and}
  \bibinfo{author}{\bibfnamefont{E.}~\bibnamefont{Artacho}},
  \bibinfo{journal}{Phys. Rev. B} \textbf{\bibinfo{volume}{66}},
  \bibinfo{pages}{205101} (\bibinfo{year}{2002}).

\bibitem[{\citenamefont{Murnaghan}(1944)}]{Murnaghan}
\bibinfo{author}{\bibfnamefont{F.~D.} \bibnamefont{Murnaghan}},
  \bibinfo{journal}{Proc. Natl. Acad. Sci.} \textbf{\bibinfo{volume}{30}},
  \bibinfo{pages}{244} (\bibinfo{year}{1944}).

\bibitem[{\citenamefont{Madelung et~al.}(2001)\citenamefont{Madelung,
  R$\mbox{\"o}$ssler, and Schulz}}]{LandoltBornstein}
\bibinfo{author}{\bibfnamefont{O.}~\bibnamefont{Madelung}},
  \bibinfo{author}{\bibfnamefont{U.}~\bibnamefont{R$\mbox{\"o}$ssler}},
  \bibnamefont{and} \bibinfo{author}{\bibfnamefont{M.}~\bibnamefont{Schulz}},
  \emph{\bibinfo{title}{Group IV Elements, IV-IV and III-V Compounds Part a}},
  vol. \bibinfo{volume}{41A1a} (\bibinfo{publisher}{Springer-Verlag -- The
  Landolt-B$\mbox{\"o}$rnstein Database}, \bibinfo{year}{2001}).

\bibitem[{\citenamefont{Hamann}(1996)}]{HamannPRL96}
\bibinfo{author}{\bibfnamefont{D.~R.} \bibnamefont{Hamann}},
  \bibinfo{journal}{Phys. Rev. Lett.} \textbf{\bibinfo{volume}{76}},
  \bibinfo{pages}{660} (\bibinfo{year}{1996}).

\bibitem[{\citenamefont{Watkins}(1986)}]{watkins1986}
\bibinfo{author}{\bibfnamefont{G.~D.} \bibnamefont{Watkins}},
  \emph{\bibinfo{title}{Deep Centers in Semiconductors}}
  (\bibinfo{publisher}{Gordon and Breach Science Publishers},
  \bibinfo{address}{New York}, \bibinfo{year}{1986}),
  chap.~\bibinfo{chapter}{3}, p. \bibinfo{pages}{147}.

\bibitem[{\citenamefont{$\mbox{\"O}$g$\mbox{\"u}$t and
  Chelikowsky}(2001)}]{serdar2001}
\bibinfo{author}{\bibfnamefont{S.}~\bibnamefont{$\mbox{\"O}$g$\mbox{\"u}$t}}
  \bibnamefont{and} \bibinfo{author}{\bibfnamefont{J.~R.}
  \bibnamefont{Chelikowsky}}, \bibinfo{journal}{Phys. Rev. B}
  \textbf{\bibinfo{volume}{64}}, \bibinfo{pages}{245206}
  (\bibinfo{year}{2001}).

\end{thebibliography}

\end{document}